\documentstyle[aps,12pt,manuscript]{revtex}
\begin{document}
\preprint{INJE--TP--95--7 }
\def\overlay#1#2{\setbox0=\hbox{#1}\setbox1=\hbox to \wd0{\hss #2\hss}#1%
\hskip -2\wd0\copy1}

\title{ Unstable two-dimensional  extremal black holes}

\author{ H.W. Lee and Y. S. Myung }
\address{Department of Physics, Inje University, Kimhae 621-749, Korea}
\author{Jin Young Kim}
\address{Division of Basic Science, Dongseo University, Pusan 616-010, Korea}

\maketitle
\vskip 1.5in

\begin{abstract}
We obtain the $\epsilon<2$ new extremal ground states of  a two-dimensional
(2D)
 charged black hole where $\epsilon$ is the dilaton coupling parameter  for
the Maxwell term.
The stability analysis is carried out for all these extremal black holes.
It is found that the shape of potentials to an on-coming tachyon (as a
spectator) take all
barrier-well types. These provide the  bound state solutions, which
imply  that they  are unstable. We conclude that the 2D, $\epsilon<2$
extremal black holes
should  not be considered as a toy model for the stable endpoint of the
Hawking evaporation.

\end{abstract}
\vskip .5in
PACS number(s) : 04.70.Bw, 04.60.Kz, 11.25.Db

\newpage
Recently the extremal black holes have received much attention.  Extremal
black holes provide
a simple laboratory in which to investigate the quantum aspects of black
hole [1]. One of the crucial
features of these extremal black holes is that the Hawking temperature
vanishes.
The black hole with $M>Q$  will evaporate down to its extremal $M=Q$ state.
Thus the extremal black hole may play  the role of
the stable endpoint for the Hawking evaporation. It has been also proposed
that although
the extremal black hole has nonzero area, it has zero entropy [2]. This is
because the extremal case is
distinct topologically  from the nonextremal one.

It is very important to investigate the stability of the extremal black
holes, which  is essential
to establish their
physical existence. It has been shown that all 4D extremal  charged black
holes  with the
coupling parameter ($a$)  are
shown to be classically  stable [3]. The $a=0$ case corresponds to
the Reissner-Nordstr\"om black hole.  Since all potentials are positive
definite outside of the
outer horizon, one easily infer  the stability of 4D extremal  charged
black holes  using the same
argument as employed by Chandrasekhar [4]. More recently,
it is shown that the 2D  extremal black holes are unstable [5].

In this Letter, we consider the two-dimensional dilaton gravity coupled to
Maxwell and tachyon fields.
The relevant coupling (parametrized by  $\epsilon$) between the dilaton and
Maxwell field
is included to obtain the general black hole. This may be considered as a
two-dimensional counterpart
of the 4D dilaton gravity with the parameter $a$.
This coupling allows us to study the new  $ 0<\epsilon< 2$ and $\epsilon
<0$ extremal black holes.
The $\epsilon=0$ case corresponds to the 2D electrically charged extremal
black hole.  Our main task is to show whether these new extremal black
holes are stable or not.
As a  compact criterion, the  black bole is unstable if there exists
a well-type potential to the on-coming physical waves [6,7].
This is so because, in solving the Schr\"odinger equation,
the potential well always allows the bound state as well as scattering state.
The former shows up as an imaginary frequency mode, leading to an
exponentially growing mode.
This indicates that the black hole is unstable.

We  start with  two-dimensional dilaton ($\Phi$) gravity  conformally
coupled to Maxwell ($F_
{\mu\nu}$) and tachyon ($T$) fields [8]
\begin{equation}
S = \int d^2 x \sqrt{-G} e^{-2\Phi}
   \big \{ R + 4 (\nabla \Phi)^2 +\alpha^2 - {1 \over 2}e^{\epsilon \Phi} F^2 -
{1 \over 2} (\nabla T)^2  +T^2\big \}.
\end{equation}
The above action with $\epsilon=0$ corresponds to the
heterotic string. The tachyon is introduced to study the properties of
black holes in a simple way.
Setting $\alpha^2 = 8$ and  after deriving equations of motion, we take the
transformation
\begin{equation}
-2\Phi \rightarrow \Phi,~~~ T \rightarrow \sqrt 2 T, ~~~-R  \rightarrow R.
\end{equation}
Then the equations of motion become
\begin{eqnarray}
&&R_{\mu\nu} + \nabla_\mu \nabla_\nu \Phi  + \nabla_\mu T \nabla_\nu T
+{ 4 - \epsilon \over 4}e^{-\epsilon \Phi/ 2}  F_{\mu\rho}F_{\nu}^{~\rho} =
0,  \\
&& \nabla^2 \Phi + (\nabla \Phi)^2  - {1 \over 2}e^{-\epsilon \Phi/ 2} F^2
- 2 T^2 - 8 = 0,  \\
&&\nabla_\mu F^{\mu \nu} + {2- \epsilon \over 2} (\nabla_\mu \Phi) F^{\mu
\nu} = 0,   \\
&&\nabla^2 T + \nabla \Phi \cdot \nabla T + 2 T = 0.
\end{eqnarray}

The general solution to the above equations is given by

\begin{equation}
\bar \Phi = 2 \sqrt 2 r,~~~ \bar F_{tr} = Q_E e^{-(2-\epsilon) \sqrt 2
r},~~~ \bar T = 0,
{}~~~ \bar G_{\mu\nu} =
 \left(  \begin{array}{cc} - f & 0  \\
                            0 & f^{-1}   \end{array}   \right)
\end{equation}
with
\begin{equation}
f = 1 -  {M \over \sqrt 2}e^{- 2 \sqrt 2 r} + {Q^2_E \over 4(2-
\epsilon)}e^{- (4-\epsilon) \sqrt 2 r},
\end{equation}
where $M$ and $Q_E$ are the mass and electric charge of the black hole,
respectively.
Note that from the requirement of $\bar F(r\to \infty) \to 0$ and $f(r\to
\infty) \to 1$, we have
the important constraint : $\epsilon <2$.
Hereafter we take $ M=\sqrt2$ for convenience.    In general, from $f=0$ we
can obtain
two roots ($r_{\pm}$) where $r_{+}(r_{-})$ correspond to the event (Cauchy)
horizons.
The extremal black hole may provide a toy model to investigate the late stages
of Hawking evaporation. We are mainly interested  in the extremal limit
(multiple root: $r_+=r_- \equiv r_o$)
of the charged black holes.
The multiple root is obtained when $f(r_o) =0$ and $f^\prime (r_o) = 0$, in
which case the square of charge is
$Q_E^2= 8({2-\epsilon \over 4-\epsilon})^{(4-\epsilon)/2}$.
Here the prime $(\prime)$ denotes the derivative with respect to $r$.
The extremal horizon is located at
\begin{equation}
 r_o(\epsilon)= - {1 \over 2 \sqrt 2} \log ({ 4-\epsilon \over 2-\epsilon}).
\end{equation}
The explicit form of the extremal $f$ is
\begin{equation}
f_e(r,\epsilon) = 1 - e^{- 2 \sqrt 2 r} + {2 \over (2- \epsilon)} \big (
{2-\epsilon \over 4-\epsilon} \big)
^{(4-\epsilon)/2}e^{- (4-\epsilon) \sqrt 2 r}.
\end{equation}
Fig. 1 shows  the multiple roots of $f_e= 0$ at $r_o(\epsilon)
= -1.076, -0.299,$ and $-0.119$ for $\epsilon $= 1.9, 0.5, and $-3$
respectively.

Now let us study whether these extremal ground states are stable or not.
We introduce small perturbation fields  around
the background solution as [5]
\begin{eqnarray}
&&F_{tr} = \bar F_{tr} + {\cal F}_{tr} = \bar F_{tr} [1 - {{\cal F}(r,t)
\over Q_E}],        \\
&&\Phi = \bar \Phi + \phi(r,t),                       \\
&&G_{\mu\nu} = \bar G_{\mu\nu} + h_{\mu\nu}  = \bar G_{\mu\nu} [1 - h
(r,t)],     \\
&&T = \exp (-{\bar \Phi \over 2}) [ 0 + t (r,t) ].
\end{eqnarray}
 One has to linearize (3)-(6) in order to obtain the equations governing
the perturbations.
However, the stability should be based on the physical degrees of freedom.
It is  thus important to check whether the graviton ($h$),  dilaton
($\phi$), Maxwell
mode (${\cal F}$)  and tachyon ($t$) are  physically propagating modes
in the 2D charged black hole background.
We review the conventional counting of degrees of freedom.
The number of degrees of freedom for the gravitational field ($h_{\mu\nu}$) in
$D$-dimensions is $(1/2) D (D -3)$.  For a Schwarzschild black hole,
we obtain two degrees of freedom. These correspond to the Regge-Wheeler
mode for odd-parity perturbation
and Zerilli mode for even-parity perturbation [4].  We have $-1$ for $D=2$.
This means that in
two dimensions
the contribution of the graviton is equal and opposite to that of a
spinless particle (dilaton).
The graviton-dilaton modes ($h+\phi, h-\phi$) are gauge degrees of freedom
and thus turn out to be
nonpropagating modes[6].
In addition, the Maxwell field has $D-2$ physical degrees of freedom.
The Maxwell field has no physical degrees of freedom for $D=2$.
Since all these fields are  nonpropagating modes, we will not consider
equations (3)-(5).
However, the tachyon as a spectator is a physically propagating mode.
This is introduced  to illustrate many of the qualitative results about the
2D charged black hole
in a simpler context.
Its linearized equation is
\begin{equation}
f^2 t''  + ff't' - [\sqrt 2 f f' -2 f(1-f)]t  - \partial_t^2 t = 0.
\end{equation}
To study the stability, we transform the above equation
 into  one-dimensional Schr\"odinger equation.
Introducing a tortoise coordinate
$$r\to r^* \equiv g(r),$$
(15) can be rewritten as
\begin{equation}
f^2 g'^2 {\partial^2 \over \partial r^{*2}} t  + f \{ f g'' +  f' g'\}
{\partial \over \partial r^* }t - \{\sqrt 2 ff' - 2 f (1 - f)\}t
 - {\partial^2 \over \partial t^2} t = 0,
\end{equation}
Requiring that the coefficient of the linear derivative vanish, one finds
the relation
\begin{equation}
g' =  {1 \over f}.
\end{equation}
Assuming $t( r^*,t ) \sim \tilde t ( r^* ) e^{i\omega t}$,
one can cast (16) into the Schr\"odinger equation

\begin{equation}
\{ {d^2 \over dr^{*2}} + \omega^2 - V(r)\} \tilde t = 0,
\end{equation}
where the effective potential $V(r)$  is given by
\begin{equation}
V(r) = f\{\sqrt 2 f' + 2  (f - 1)\}.
\end{equation}
We are interested only in the extremal black holes.
The potentials surrounding the extremal black holes are given by
\begin{equation}
V_e(r,\epsilon)  = 2 e^{- 2 \sqrt 2 r} f_e \{ 1- 2({3-\epsilon \over
2-\epsilon})
({2-\epsilon \over 4-\epsilon})^{(4-\epsilon)/2}e^{- (2-\epsilon) \sqrt 2 r}\}.
\end{equation}
After a concrete analysis, one  finds the barrier-well type potentials for
$\epsilon <2$.
For examples, Fig. 2 shows the shapes of potentials for $\epsilon $= 1.9,
0.5, and $-3$.

Now let us translate the potential $V_e(r,\epsilon)$ into $V_e(r^*, \epsilon)$.
With $f_e$ in (10), one finds the form of $r^*= \int^r {dr \over f_e}$
\begin{equation}
r^*= r + {1 \over 2 \sqrt 2(4-\epsilon)} \log |f_e|
- {2-\epsilon \over 2 \sqrt 2 (4- \epsilon)} \int ^y {dy \over { 1- y +A
y^{1+ B}}}
\end{equation}
with $y= e^{-2 \sqrt 2 r}, A= {2 \over (2- \epsilon)} [{2-\epsilon \over
4-\epsilon}]
^{(4-\epsilon)/2},$ and $B = 1- {\epsilon \over 2}$.
Since both the forms of $V_e(r,\epsilon)$ and $r^*$ are very complicated,
we are far from obtaining
the exact form of $V_e(r^*,\epsilon)$. At this point, it is crucial to
obtain  the approximate forms
to $V_e(r^*,\epsilon)$  near the both ends.
In the asymptotically flat region  one finds from (21) that $r^* \simeq r$.
(20) takes
the asymptotic form
\begin{equation}
V_{r*\to \infty} \simeq 2 \exp( -2 \sqrt 2 r^*),
\end{equation}
which is independent of $\epsilon$.
On the other hand, near the horizon ($r=r_o$) one has
\begin{equation}
r^* \simeq - { 2 \over \sqrt 2 (2-\epsilon)} { 1 \over ({4-\epsilon \over
2-\epsilon}-
  e^{ - 2 \sqrt 2 r})}.
\end{equation}
Approaching the horizon $(r\to r_o, r^* \to -\infty)$, the potential takes
the form
\begin{equation}
V_{r*\to -\infty} \simeq - {1 \over (4-\epsilon)}{ 1 \over r^{*2}}.
\end{equation}
Using (22) and (24) one can construct the approximate form $V_{app}(r^*,
\epsilon)$ (Fig. 3).
This is also a  barrier-well which is  localized at the origin of $r^*$.

Our stability analysis is  based on the equation
\begin{equation}
\{ {d^2 \over dr^{*2}} + \omega^2 - V_{app}(r^*,\epsilon)\} \tilde t = 0.
\end{equation}
As is well known, two kinds of solutions to the Schr\"odinger equation
correspond to
 the bound and scattering states. In our case $V_{app}(r^*)$ admits  two
solutions  depending on
the signs of the energy ($E=\omega^2$):
(i) For $E>0 (\omega=$ real), the asymptotic solution for $\tilde t$ is given
by
$\tilde t_\infty  =  \exp(i \omega r^*)  + R \exp(- i \omega r^*), r^*  \to
\infty $ and
$\tilde t_{EH} = T\exp( i\omega r^*), r^*  \to - \infty $.
Here $R$ and $T$ are the scattering amplitudes of two waves which are
reflected and transmitted by the potential $V_{app}(r^*,\epsilon)$, when a
tachyonic wave of unit
amplitude with the frequency $\omega$ is incident on the black hole from
infinity.
\noindent $(ii)$ For $E<0 (\omega =-i \alpha$, $\alpha$ is positive and real),
 we have the bound state.
Eq. (25) is given by

\begin{equation}
{d^2 \over d r^{*2}}\tilde t  =  (\alpha^2 + V_{app}(r^*)) \tilde t.
\end{equation}
The asymptotic solution is $\tilde t_\infty  \sim  \exp(\pm \alpha r^*),
r^*  \to \infty $
and $\tilde t_{EH}    \sim  \exp(\pm \alpha r^*), r^*  \to - \infty $.
To ensure that the perturbation falls off to zero for large $r^*$, we choose
$\tilde t_\infty \sim \exp (-\alpha r^*)$.  In the case of $\tilde t_{EH}$,
the solution
$\exp (\alpha r^*)$ goes to zero as $r^* \to - \infty$.
Now let us observe whether or not $\tilde t_{EH} \sim \exp (\alpha r^*)$
can be matched
to $\tilde t_\infty \sim \exp (-\alpha r^*)$.
Assuming $\tilde t$ to be positive,
the sign of $d^2 \tilde t / dr^{*2}$
can be changed from $+$ to $-$ as  $r^*$
goes from $\infty$ to $-\infty$.
If we are to connect $\tilde t_{EH}$ at one end to a decreasing solution
$\tilde t_\infty$
at the other, there must be a point ($d^2\tilde t/ dr^{*2}<0$,
$d \tilde t/dr^*=0$) at which the signs of $\tilde t$ and
$d^2\tilde t/dr^{*2}$ are opposite : this is  compatible with  the shape of
$V_{app}(r^*,\epsilon)$ in Fig. 3. Thus it is possible for
$\tilde t_{EH}$ to be connected to $\tilde t_\infty$ smoothly.  Therefore a
bound state solution
is given by

\begin{eqnarray}
\tilde t_\infty & \sim & \exp(- \alpha r^*),~~~~~~~~ ( r^*  \to \infty )     \\
\tilde t_{EH}   & \sim & \exp( \alpha r^*), ~~~~~~~~ ( r^*  \to - \infty ).
\end{eqnarray}
This is a regular solution everywhere in space at the initial time $t=0$.
It is well-known that in quantum mechanics, the bound state solution is
always allowed if
there exists a potential well.
The time evolution of the solution with $\omega=-i\alpha$ implies
$t_\infty (r^*, t)=\tilde t_\infty(r^*)\exp(-i\omega t) \sim \exp (-\alpha
r^*) \exp (\alpha t)$ and
$t_{EH} (r^*,t)=\tilde t_{EH}(r^*) \exp(-i\omega t) \sim \exp (\alpha r^*)
\exp (\alpha t)$.
This means that there exists an exponentially growing mode with time.
Therefore, the $\epsilon<2$ extremal ground states  are
classically unstable. The origin of this instability comes from  the
barrier-well potentials.
These potentials  appear  from  all $\epsilon<2$ extremal black holes.

We conclude that the 2D, $\epsilon<2$  extremal  black holes cannot be
considered as a toy model for the stable endpoint of the Hawking evaporation.

\acknowledgments

This work was supported in part by Nondirected Research Fund, Korea
Research Foundation,1994
and by Korea Science and Engineering Foundation, 94-1400-04-01-3.

\figure{Fig. 1: Three graphs of extremal $f_e$ for $\epsilon =1.9$ (dashed
line : $-$-$-$-),
0.5 (dotted line  :- - - -), and $-3$ (solid line : ---).
The corresponding multiple roots (extremal horizons) are located at
$r_o = -1.076, -0.299$, and  $-0.119$ respectively.

Fig. 2: Three graphs of extremal potentials ($V_e(r,\epsilon)$) for
$\epsilon = 1.9$ (dashed line : $-$-$-$-),
0.5 (dotted line  :- - - -), and $- 3$ (solid line : ---). The potentials
are zero at $r_o(\epsilon)$
and  all barrier-well types outside $r_o(\epsilon)$ .

 Fig.3 : The approximate  potential ($V_{app}(r^*,\epsilon)$).
The aymptotically flat region is at $r^* = \infty$.
This also takes a  barrier-well type. This is localized at $r^*=0$, falls
to zero
exponentially as $r^* \to \infty$ and inverse-squarely as $r^* \to -\infty$
(solid lines).
 The dotted line is used to connect two boundaries.}
\end{document}